\documentclass[11pt]{article}
\usepackage[font=small]{caption}
\usepackage[letterpaper, margin=1in]{geometry}
\usepackage{setspace}
\usepackage{graphicx}
\usepackage{amsmath}
\usepackage{amssymb}
\usepackage{booktabs}
\usepackage{hyperref}
\usepackage{xcolor}
\usepackage{listings}
\usepackage[numbers]{natbib}
\usepackage{comment}
\usepackage{siunitx}
\usepackage[margin=5pt,font=footnotesize]{subfig}
\usepackage{enumitem}
\usepackage{makecell}

\hypersetup{
    colorlinks,
    linkcolor={red!50!black},
    citecolor={blue!50!black},
    urlcolor={blue!80!black}
}

\begin{document}

\title{All-in-One Deep Learning Framework for MR Image Reconstruction}

\author{Geunu Jeong\textsuperscript{*}, Hyeonsoo Kim, Joonyoung Yang, Kyungeun Jang, Jeewook Kim}

\date{}
\maketitle
\vspace{-0.8cm}
\begin{center}
AIRS Medical, Seoul, Korea \\

\medskip
* Correspondence: \href{mailto:jeong.geunu@airsmed.com}{jeong.geunu@airsmed.com}
\end{center}

\begin{abstract}
\noindent
\textbf{Purpose:} This research introduces a novel, \textit{all-in-one} deep learning framework for MR image reconstruction, enabling a single model to enhance image quality across multiple aspects of $k$-space sampling and to be effective across a wide range of clinical and technical scenarios. \\
\textbf{Methods:} 150,000 MR raw data of various pulse sequences and anatomical regions from three vendors were collected. Multi-dimensional degradation was applied to raw $k$-space data to generate training input. This process involved a combined application of noise addition and multiple patterns of undersampling (uniform, random, $k_{max}$, partial Fourier, elliptical), with each method being applied across a range of factors to cover extensive sampling scenarios. Contextual data, including scan parameter information, were prepared to serve as auxiliary input to address the challenges posed by the unique learning task for each training pair, which arise from the varied degradation scenarios. The expected noise reduction factor for each training pair was mathematically derived and used as additional contextual data for tunable denoising. The U-Net was modified to include an additional pathway for integrating contextual data. \\
\textbf{Results:} Seven performance evaluations were conducted through visual comparisons and quantitative analysis. 1) A series of deep learning reconstructions (DLRs) was applied to the same image, each adding a new dimension of improvement - starting with noise reduction, then adding frequency $k_{max}$, phase partial Fourier, phase $k_{max}$, to slice $k_{max}$. Incremental enhancement along each added dimension was demonstrated, confirming simultaneous multi-dimensional improvements. 2) Measured noise levels in DLR images corresponded to the five different applied noise reduction factors, showcasing the accuracy of tunable denoising. 3) A relative edge sharpness of approximately 2.5 between original and DLR images, indicating effective super-resolution, was achieved in each of the three encoding directions, independent of the noise reduction factor. 4) DLR images showed reduced truncation and intravoxel dephasing artifacts, which are prominent at lower resolutions, attributed to slice-directional super-resolution. 5) Images obtained from eight different sets of scan parameters that adjusted sampling and reconstruction pipeline, along with 6) images from three unseen vendors, demonstrated significant quality improvement after DLR, highlighting broad compatibility. 7) Image pairs from standard and fast protocols across four anatomical regions were acquired, with the fast images undergoing DLR. The DLR fast images exhibited superior quality compared to the standard images, demonstrating the feasibility of reducing scan times. \\
\textbf{Conclusions:} The proposed model enhances image quality in a multi-dimensional manner and offers versatility. \\
\textbf{Clinical Relevance/Application:} The proposed model is compatible with a broad spectrum of scenarios, including various vendors, pulse sequences, scan parameters, and anatomical regions. Its DICOM-based operation particularly enhances its applicability for real-world applications. 
\end{abstract}

\section{Introduction}
Magnetic Resonance Imaging (MRI) stands out among clinical imaging modalities due to its excellent soft tissue contrast and the absence of ionizing radiation. However, achieving sufficient diagnostic image quality often requires lengthy scan times, which can reduce patient throughput and cause discomfort. Consequently, most clinical applications face the dilemma of having to choose between prolonged scan times and compromised image quality. This dilemma necessitates enhancing the quality of images obtained from shorter scan times to effectively resolve the issue.

Research into deep learning (DL)-based MR image reconstruction has become increasingly active in recent years to address this persistent need \cite{hammernik2018learning, wang2016accelerating}. Notably, the 2019 and 2020 fastMRI challenges, hosted by Facebook AI Research and NYU Langone Health, were competitions aimed at tackling this issue \cite{knoll2020advancing, muckley2021results}. The majority of these efforts have focused on reconstructing fully-sampled data from highly undersampled data, either uniformly or randomly \cite{muckley2021results, hammernik2021systematic, clifford2022artificial, pezzotti2020adaptive}. These approaches aim to enhance the quality of images obtained from shorter scan times so that they match those obtained from longer scan times, thereby challenging the conventional trade-off between scan time and image quality.

However, a significant limitation of most existing frameworks lies in their focus on a single dimension\footnote{In this paper, the term "dimension" in the context of image quality does not refer to geometric dimensions but signifies multiple aspects of $k$-space sampling.} in improving image quality. Their methodology for generating training pairs, involving uniform or random undersampling, fundamentally restricts the scope of potential image quality improvements to this specific dimension. Yet, the range of dimensions that determine image quality extends well beyond this limited focus, including key $k$-space sampling defining factors such as $k_{max}$, ${\Delta k}$, partial Fourier factor, elliptical sampling factor, and noise variance of samples. Uniform or random undersampling pertains solely to one of these dimensions, ${\Delta k}$. While these dimensions share the common effect of influencing an image's noisy or blurry appearance, each has a distinct mathematical impact on image quality. Therefore, a methodology that considers all these diverse dimensions for creating training pairs could potentially enable multi-dimensional improvement in image quality. This approach also opens avenues for further reducing scan times across more dimensions. Furthermore, even when employing uniform or random undersampling for image acceleration, as with previous approaches, there's a potential not just for restoring quality to match fully-sampled data but also for surpassing it across additional dimensions.

Another limitation of many previous approaches is their restricted coverage of scenarios in training input simulations, which could potentially narrow the scope of compatible scan parameters. The performance of a DL model is typically enhanced when the inference input is similar to the inputs encompassed within the training dataset. This similarity can be defined across multiple scan parameters, and the more these features align between the training and inference inputs, the higher the likelihood of achieving optimal performance. Given these considerations, it becomes crucial to cover a wide range of input scenarios, including diverse combinations of multiple scan parameters, to ensure the model's broad compatibility. However, most previous methods focused solely on the acceleration factor when determining input similarity and simulating scenarios. Even if the acceleration factor of the inference input matches that of the training input, a lack of similarity in other $k$-space sampling defining factors, such as $k_{max}$, can reduce the overall alignment between inputs. Furthermore, within a Digital Imaging and Communications in Medicine (DICOM)-based DL approach, not only the scan parameters influencing the $k$-space sampling but also those affecting the reconstruction pipeline may have to be considered in defining input similarity. This reconstruction pipeline consists of multiple steps, such as zero-padding interpolation, surface coil intensity normalization, and geometric distortion correction, each offering various methods and parameters. The practical utility of a model, in terms of how extensively it can be used, hinges on its ability to maintain optimal performance across a diverse range of imaging scenarios. This underscores the importance of comprehensive training on extensive combinations of scan parameters that define the $k$-space sampling and the reconstruction pipeline.

In this work, we introduce a DICOM-based, \textit{all-in-one} DL framework, enabling a single model to enhance image quality across multiple aspects of $k$-space sampling and to be effective across a wide range of clinical and technical scenarios. This MR image reconstruction algorithm serves as the core of SwiftMR\textsuperscript{TM} (AIRS Medical, Seoul, Korea), which is FDA-cleared, CE-certified, and commercially available. We first detail the comprehensive development process of the model, including data collection, training pair preparation, model architecture design, and DICOM inference. We then assess the model's capability to enhance image quality in a multi-dimensional manner, specifically across various aspects of $k$-space sampling. Subsequently, we evaluate several features of the multi-dimensional enhancement: the accuracy of tunable denoising, the effectiveness of super-resolution in each encoding direction, and the reduction of artifacts that become more prominent at lower spatial resolutions. Additionally, we assess its compatibility with various scan parameter sets and its generalizability across scanner vendors not seen during training. Finally, we present specific cases demonstrating the model's utility in reducing scan time across anatomical regions in conjunction with protocol optimization.

\section{Methods}
In this section, we describe the entire model development process, organized into four main steps: (1) data collection, (2) preparation of training pairs, (3) design of the model architecture and the training details, and (4) DICOM inference details.

\subsection{Data collection}
MR raw data were collected from several medical centers with approval from the Institutional Review Board, consisting of approximately 100,000 2D pulse sequence data and 50,000 3D pulse sequence data. This collection includes a range of scanner models and field strengths from three major vendors: Siemens Healthineers (Erlangen, Germany), GE Healthcare (Waukesha, WI, USA), and Philips Healthcare (Best, The Netherlands). It spans a broad spectrum of contrast weightings, pulse sequences, and scan parameters and covers virtually all anatomical regions and a variety of pathologies. This comprehensive approach aims to ensure clinical and technical diversity, thereby enhancing the model's applicability across various imaging scenarios. Additionally, the collection includes data that significantly exceed the quality of clinical standards to achieve superior model performance. The data were obtained through dedicated sessions with volunteers who provided informed consent, with each imaging sequence lasting about 15 minutes.

\subsection{Training pair preparation}
A hybrid approach was employed for the preparation of training data: Training pairs originated from raw $k$-space data but were presented as conventionally reconstructed images. This method of deriving training pairs from raw $k$-space data, rather than DICOM image data, was for a closer approximation to a wide variety of real acquisition scenarios. Furthermore, configuring the training pairs as conventionally reconstructed images, as opposed to $k$-space data, was to enable the model to operate on a DICOM-based framework.

The process for generating training pairs is summarized as follows: The target $k$-space was designated as the raw $k$-space data itself. For the input, raw $k$-space data underwent multi-dimensional degradation, spanning a range of factors and levels, resulting in the generation of input $k$-space. Subsequently, both input and target $k$-space were conventionally reconstructed into image pairs under a variety of reconstruction scenarios.

The methods employed to degrade the raw $k$-space data for input $k$-space generation include the following, with Figure~\ref{fig:kspace_degradation} illustrating an example of a 3D pulse sequence:

\begin{figure}[htb!]
    \centering
    \includegraphics[width=1\linewidth]{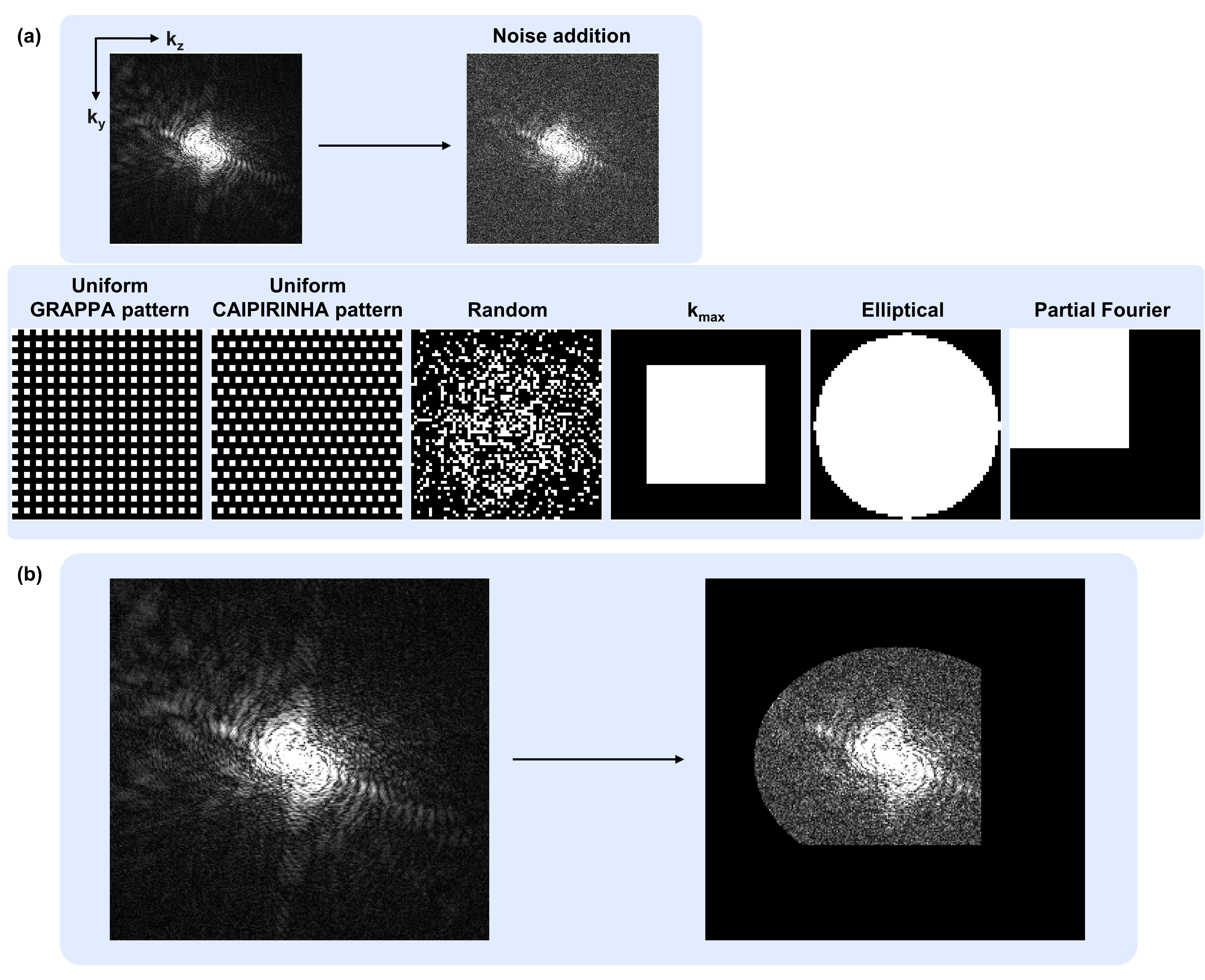}
    \caption{Schematic representation of raw $k$-space degradation methods for input $k$-space generation. For simplicity, the $k_{x}$ axis is omitted, illustrating only the $k_{y}-k_{z}$ plane. (a) Illustrates the addition of Gaussian noise and various undersampling patterns used to degrade raw $k$-space data, which are combined to generate the input $k$-space. (b) Presents one specific example featuring a combination of noise addition, $k_{max}$ undersampling, elliptical undersampling, and partial Fourier undersampling. GRAPPA, GeneRalized Autocalibrating Partially Parallel Acquisitions; CAIPIRINHA, Controlled Aliasing in Parallel Imaging Results in Higher Acceleration. 
}
    \label{fig:kspace_degradation}
\end{figure}

\begin{itemize}[noitemsep]
\item Adding Gaussian noise to samples.
\item Uniform pattern undersampling.
\item Random pattern undersampling.
\item $k_{max}$ undersampling across all encoding directions.
\item Elliptical undersampling.
\item Partial Fourier undersampling across all encoding directions.
\end{itemize}
Each degrading method uniquely affects the image, resulting in mathematically differentiated outcomes, although all methods aim to simulate higher noise or lower spatial resolution. For instance, adding Gaussian noise and applying uniform pattern undersampling both result in images with higher noise levels, yet the outcomes differ. Similarly, $k_{max}$ undersampling and partial Fourier undersampling both lead to lower spatial resolution images, but again, the results vary. The input $k$-space was generated by combined applications of these methods to raw $k$-space data. This strategy was designed to enable a single model to simultaneously enhance image quality across multiple dimensions.

Next, a wide range of conventional reconstruction scenarios was covered, focusing on the variations within the following steps of the reconstruction pipeline:
\begin{itemize}[noitemsep]
\item $k$-space anti-ringing filtering.
\item $k$-space zero-padding interpolation across all encoding directions.
\item Parallel imaging reconstruction.
\item Compressed sensing reconstruction.
\item Partial Fourier reconstruction.
\item Channel combination.
\item Extraction of magnitude, phase, real, and imaginary components.
\item Surface coil intensity normalization.
\item Gradient non-linearity induced geometric distortion correction.
\item Integer quantization.
\end{itemize}
Each $k$-space degradation method can be applied with a range of factors, and similarly, each reconstruction step can be implemented using several methods, each configurable with various intensities, factors, and parameters. A wide array of scenarios was simulated by spanning an extensive combination of these variations. The purpose of this comprehensive simulation is to enable a single model to be compatible and flexible across various combinations of scan parameters.

Finally, contextual data were prepared to serve as auxiliary inputs for the model. First, scan parameters that define $k$-space sampling were utilized. Each training pair, generated from different degradation scenarios, presents a distinct learning task, which poses a challenge for the model. Providing scan parameter information is intended to address this challenge. Second, the expected noise reduction factor for each simulated training pair was mathematically derived based on the principle that independent random noise is not inherently learnable, and this was then used. This context serves to inform the model of the level of noise reduction achieved, ultimately enabling adjustable denoising. These two types of contextual data were concatenated into a one-dimensional array and were subsequently fed into the model.

Approximately 1 million slices from 2D pulse sequence data and 2 million slices from 3D pulse sequence data, excluding those generated through data augmentation, were prepared for training.

\subsection{Model architecture design and training}
\begin{figure}[htb!]
    \centering
    \includegraphics[width=1\linewidth]{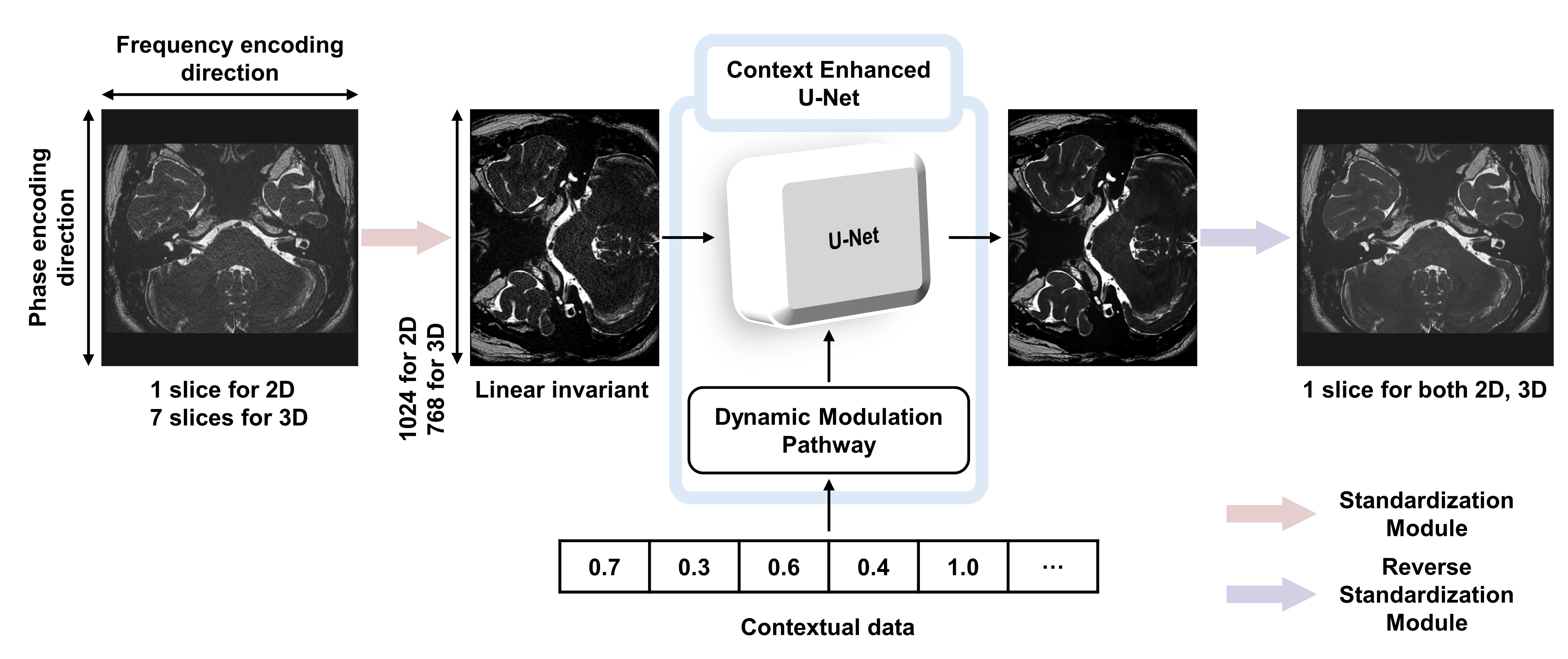}
    \caption{Proposed model architecture and overall process flow. A single slice is used as input for 2D model, and a stack of 7 adjacent slices is used for 3D model. The standardization module handles image transposing, zero cropping, in-plane interpolating and normalizing. The Context Enhanced U-Net (CE U-Net) builds on the traditional U-Net structure by incorporating a dynamic modulation pathway to integrate contextual data. The output from the CE U-Net is a single slice for both 2D and 3D. The reverse standardization module performs the reverse operations of the standardization module, except for image interpolation.}
    \label{fig:model_architecture}
\end{figure}

The DL model, designed to be fed with prepared training pairs, consists of a standardization module and a deep neural network. The standardization module is aimed at standardizing input images, thereby reducing their complexity. The utilized deep neural network, termed \textit{Context Enhanced U-Net} (\textit{CE U-Net}), is based on the traditional U-Net \cite{ronneberger2015u} but has been modified to include an additional pathway that allows for the integration of contextual data at an intermediate stage of the network. Separate networks were designed for 2D and 3D pulse sequences, with the latter specifically engineered to enable super-resolution in the direction of slice encoding. Figure~\ref{fig:model_architecture} illustrates the standardization module and the CE U-Net architecture, which are elaborated upon as follows.

The standardization module comprises four major steps. First, images are transposed as necessary to align the row direction with the phase encoding direction. This is followed by a reverse operation after the network processing, to restore them to their original orientation. Secondly, for images presenting an asymmetric displayed field of view (FOV) but adjusted to a square format through zero-padding, the unnecessary zeros are cropped out. A reverse operation is performed to return the images to their square format after the network processing. Third, the images undergo Lanczos in-plane interpolation to adjust the column size to 1024 for 2D pulse sequences and to 768 for 3D pulse sequences, ensuring consistency in image pixel spacing. Lastly, input images are normalized to become invariant to linear transformations, countering the linear rescaling of vendor DICOM. This normalization is also reversed after the network processing to retain the original dynamic range.

A Dynamic Modulation Pathway (DMP) was introduced to integrate contextual data at an intermediate stage of the U-Net, thus creating the CE U-Net. The DMP consists of fully connected layers and activation functions that process the contextual data. This processed data is then integrated into the U-Net framework, acting as convolution kernels.

The networks for 2D and 3D pulse sequence images were separately developed and trained. The 3D-specific network is designed to process a stacked input of seven adjacent slices, producing the middle slice as output. The primary rationale for this distinct architecture stems from the $k_{max}$ undersampling in the slice encoding direction for 3D pulse sequence data during the training pair preparation phase. This simulation results in sinc blurring of a wider width, thereby embedding information about the target slice into its adjacent slices. Leveraging this information makes it feasible to achieve super-resolution in the slice encoding direction.

The training details are as follows: The foundational U-Net features an initial convolutional block with 64 output channels and includes four stages of downsampling and upsampling, resulting in a total of approximately 10 million parameters. During the training phase, an L1 loss function and the ADAM optimizer \cite{kingma2014adam} were utilized to adjust the network's parameters. The learning rate was initially set at 0.001 and was subsequently reduced by a factor of 0.1 after 10 epochs, followed by an additional 3 epochs of training.

\subsection{DICOM inference}
The trained model takes DICOM data as input for inference. The image data to be processed is derived from the DICOM's pixel data tag. The phase encoding direction and displayed phase field of view, required for the standardization module, are sourced from DICOM tags. The selection between networks for 2D or 3D pulse sequences is guided by acquisition type information, also extracted from a DICOM tag. $k$-space sampling defining scan parameters, serving as contextual data, are likewise retrieved from DICOM tags. Moreover, the noise reduction factor, another contextual data, is directly input by the user, allowing for customizable denoising.

\section{Results}
In this section, we delineate the methodology employed for evaluating the performance of the proposed model, along with the subsequent results. All evaluations were carried out using DICOM data acquired directly from MR scans of volunteers, without the use of internally simulated imagery. Informed consent was obtained from all volunteers.

\subsection{Multi-dimensional image enhancement}
To assess the capability of the single model to simultaneously improve multiple dimensions of image quality, a series of deep learning reconstructions (DLRs) was applied to the same image, each incrementally including an additional dimension of improvement. This method works by adjusting the contextual data input into the model, which allows for targeted enhancements in specific dimensions of image quality. We then conducted a visual comparison between the original image and the multiple DLR images to observe the enhancement made at each additional dimension. For this purpose, a brain T2-weighted image (T2WI) was acquired using a 3D constructive interference in steady state (CISS) pulse sequence on a 1.5T MR scanner (MAGNETOM Avanto, Siemens Healthineers) with a head/neck matrix coil. The scan plane was axial, phase encoding direction was right-to-left, acquisition voxel size was 1.0mm isotropic, phase partial Fourier was 6/8, slice partial Fourier was off, elliptical sampling was off, and no acceleration technique was used. The image underwent five distinct DLRs, each focusing on specific dimensions as follows:
\begin{enumerate}[noitemsep]
    \item Noise reduction only.
    \item Adding frequency $k_{max}$ dimension.
    \item Adding phase partial Fourier dimension.
    \item Adding phase $k_{max}$ dimension.
    \item Adding slice $k_{max}$ dimension.
\end{enumerate}
The same image underwent five separate DLRs, each using different contextual data, rather than employing a sequential approach where a reconstructed image undergoes further reconstruction. A noise reduction factor of 3.0 was used for each DLR.

\begin{figure}[htb!]
    \centering
    \includegraphics[width=1\linewidth]{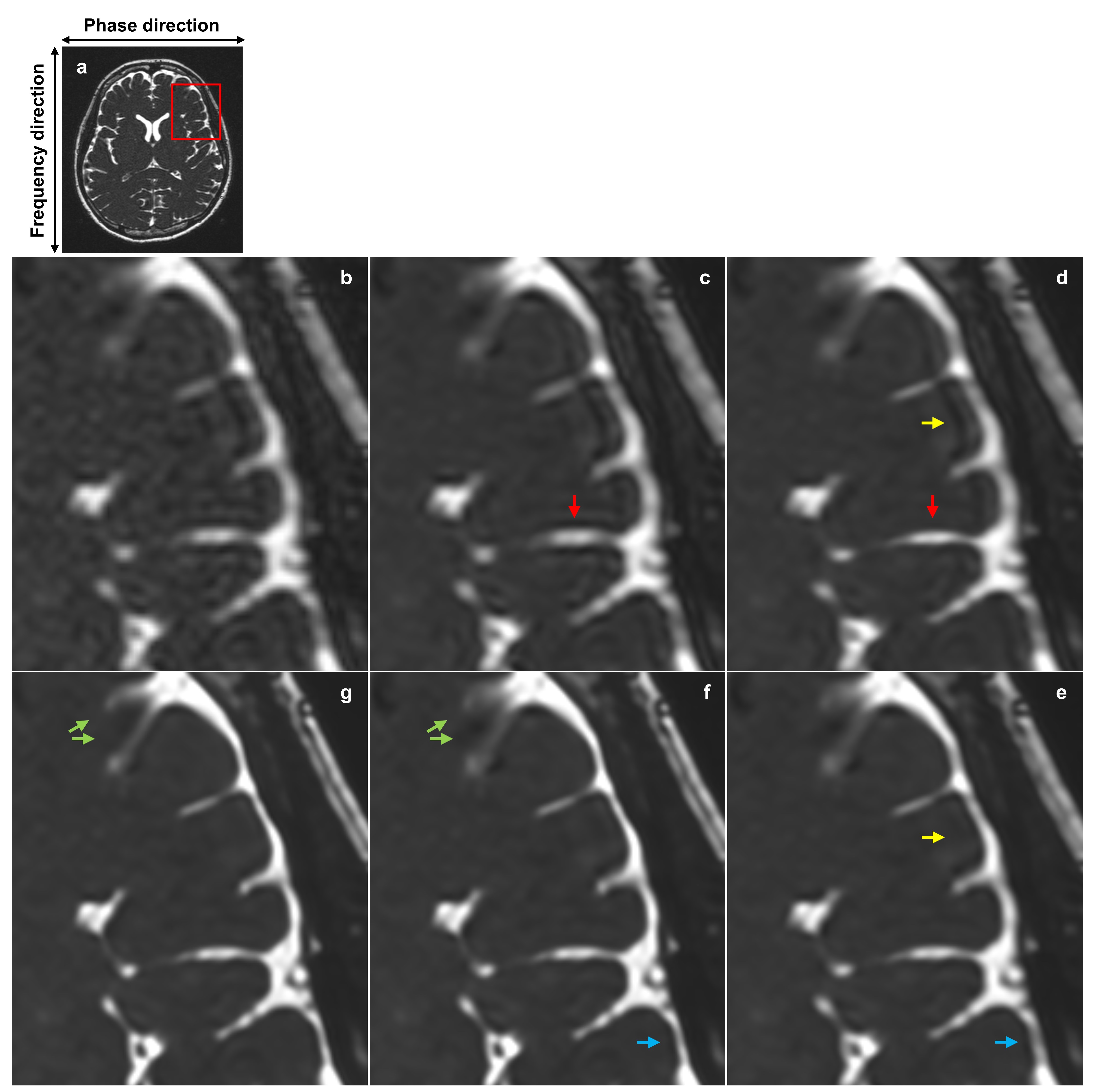}
    \caption{Visual comparison between multiple DLR images, each adding a new dimension of improvement incrementally. (a) Shows the original image. Subsequent images are magnifications of the area marked by a red rectangle in (a): (b) Magnified area of the original image. (c) Noise reduction only, demonstrating decreased noise compared to (b) without a change in perceived resolution. (d) Includes enhancement in the frequency $k_{max}$ dimension, improving perceived resolution in the frequency direction (red arrows). (e) Adds the phase partial Fourier dimension, reducing partial Fourier induced blurring (yellow arrows). (f) Incorporates the phase $k_{max}$ dimension, boosting apparent resolution in the phase direction (blue arrows). (g) Final enhancement in the slice $k_{max}$ dimension, reducing blurring due to slice directional partial volume effects (green arrows). Each of five DLRs was performed on the original image by adjusting contextual data, not by sequentially reprocessing previous DLR images. A comparison between (b) and (g) illustrates the simultaneous multi-dimensional enhancement of image quality. DLR, deep learning reconstruction.
}
    \label{fig:multi_dimensional_enhancement}
\end{figure}

Figure~\ref{fig:multi_dimensional_enhancement} illustrates the change in image as each additional dimension is incorporated. Initially, the inclusion of the noise reduction dimension results in decreased noise without altering perceived resolution. The subsequent addition of the frequency $k_{max}$ dimension enhances perceived resolution in frequency direction. Following this, including the phase partial Fourier dimension reduces partial Fourier induced blurring. Adding the phase $k_{max}$ dimension further boosts apparent resolution in phase direction. Lastly, incorporating the slice $k_{max}$ dimension diminishes image blurring caused by slice directional partial volume effects. Comparing the original image with the fifth reconstructed image, which includes all dimensions, confirms the multi-dimensional image improvement.

\subsection{Accuracy of tunable denoising}
The proposed multi-dimensional image enhancement includes denoising in an adjustable manner. To verify that the specified noise reduction factors input as contextual data result in corresponding reductions in noise levels in DLR images, we applied DLRs with different noise reduction factors to the same image and then measured the resulting noise levels. For this purpose, a brain T1-weighted image (T1WI) was acquired using a 3D turbo field echo (TFE) pulse sequence on a 1.5 T MR scanner (Multiva, Phillips Healthcare) with an 8-channel head coil. The image was DL-reconstructed with noise reduction factors of 1.0, 1.5, 2.0, 2.5, and 3.0. Mean signal and noise standard deviation for each image were calculated by placing a circular region of interest (ROI) within the pons, where the signal can be assumed to be homogeneous \cite{uetani2021preliminary}. The reason for measuring the noise level within a ROI instead of in the outside air is due to the pixel values in the outside air region of the original image being already at zero, as a result of the sensitivity-weighted channel combination. The relative noise level was calculated by taking the ratio of the noise level in the DLR image to that in the original image.

\begin{figure}[htb!]
    \centering
    \includegraphics[width=1\linewidth]{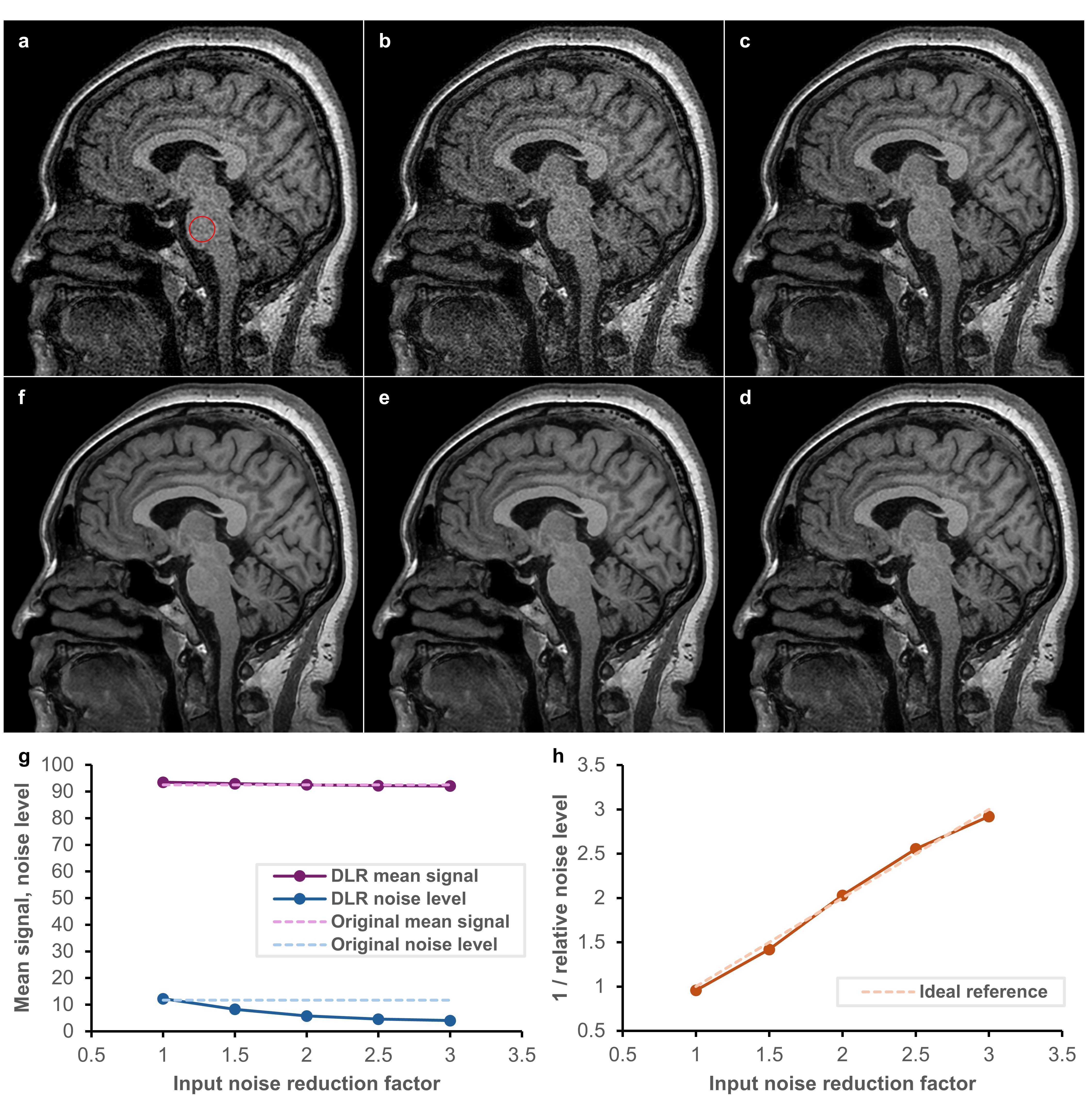}
    \caption{Mean signal and noise levels in DLR images at various input noise reduction factors. (a) Original image with a red circle indicating the region of interest used for measuring mean signal and noise levels. (b-f) DLR images at noise reduction factors of 1.0, 1.5, 2.0, 2.5, and 3.0 respectively, showing progressive noise reduction. (g) Presents the mean signal and noise levels across these factors, with mean signal remaining constant while noise level decreases. (h) Illustrates the inverse ratio of noise levels between the DLR images and original image, confirming alignment with the applied noise reduction factors. DLR, deep learning reconstruction.}
    \label{fig:tunable_denoising}
\end{figure}

Figure~\ref{fig:tunable_denoising} displays the original image and its DLR images at various noise reduction factors, along with graphs detailing measured mean signal and noise levels. As the noise reduction factor increases, a visible reduction in noise is observed. The graph shows that the mean signal intensity remains constant regardless of the noise reduction factor applied. Additionally, the achieved noise reduction levels correspond to the input noise reduction factors.

\subsection{Super-resolution in all encoding directions}
The proposed multi-dimensional image enhancement includes super-resolution in all encoding directions. To verify the effectiveness of super-resolution in each direction, we conducted a quantitative analysis comparing the edge sharpness between original and DLR images. For this purpose, a brain MR angiography (MRA) image was acquired using a 3D time of flight (TOF) pulse sequence on a 1.5 T MR scanner (MAGNETOM Altea, Siemens Healthineers) with a 20-channel head/neck coil. Maximum intensity projection (MIP) images were used to measure edge sharpness due to their abundance of linear structures. The A2 segment of the anterior cerebral artery was selected to evaluate the super-resolution effect in the frequency encoding direction in a sagittal MIP image. Meanwhile, the P1 segment of the posterior cerebral artery and the S1 segment of the superior cerebellar artery were chosen to assess the super-resolution effect in the phase and slice encoding directions, respectively, in a coronal MIP image. The method for computing relative edge sharpness was directly adopted from Lebel's approach \cite{lebel2020performance}. Line profiles parallel to each encoding direction were plotted across the selected arteries. The relative edge sharpness was then calculated as the ratio of the maximum gradient in these line profiles in the DLR images to that in the original images. Additionally, to investigate the influence of different noise reduction factors on the effectiveness of super-resolution, the source image was DL-reconstructed with variable noise reduction factors of 1.0, 1.5, 2.0, 2.5, and 3.0. DLR MIP images were generated from the DLR source image, rather than by applying DLR directly to the original MIP images. 

\begin{figure}[htb!]
    \centering
    \includegraphics[width=1\linewidth]{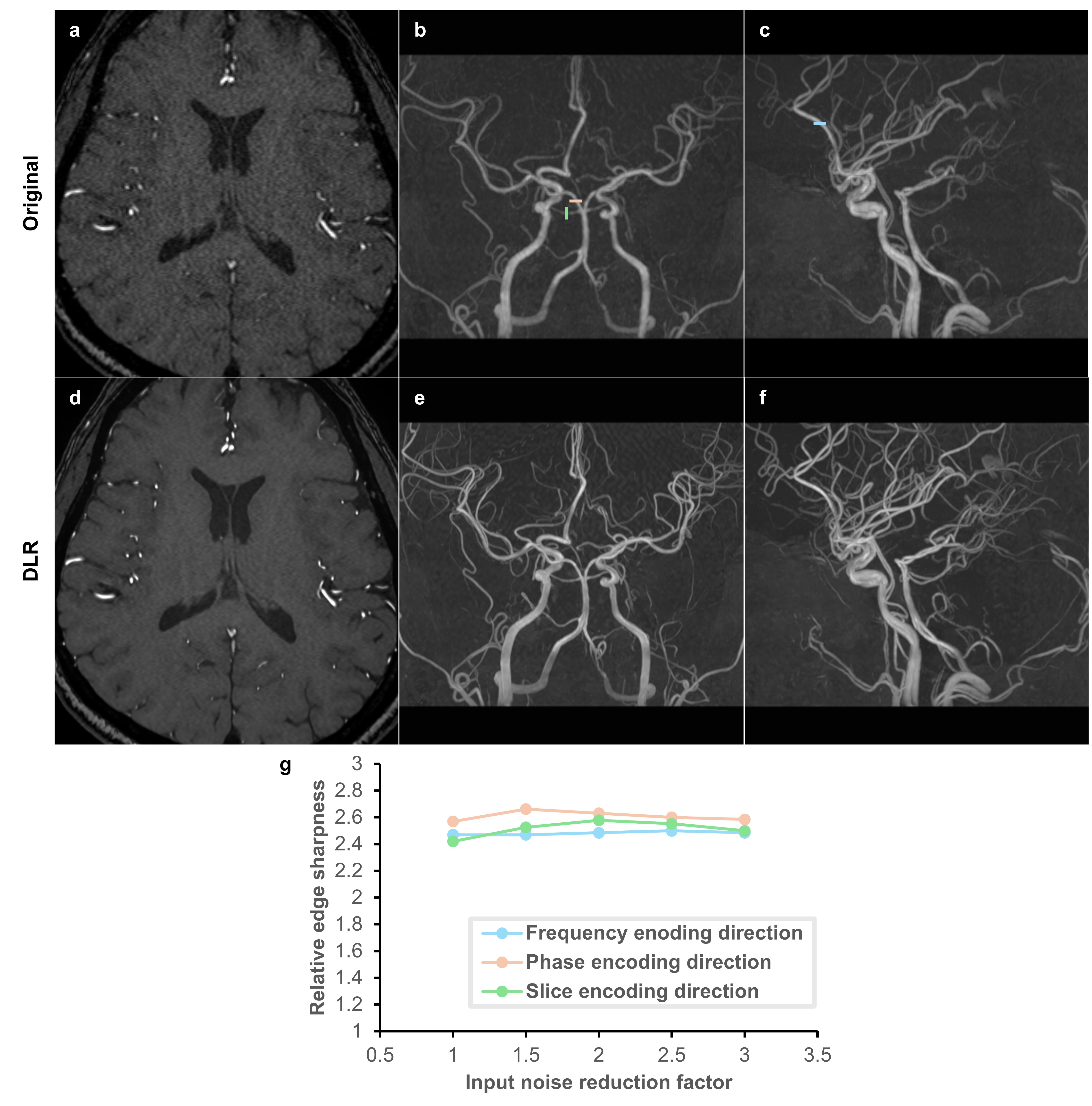}
    \caption{Relative edge sharpness between original and DLR images in three encoding directions at various noise reduction factors. (a) Original MR angiography source image. (b,c) MIP images derived from (a), with line profiles marked in blue, red, and green for the frequency, phase, and slice directions respectively, used for measuring edge sharpness. (d) DLR image of (a). (e,f) MIP images derived from (d). Comparisons between (a) and (d), (b) and (e), and (c) and (f) reveal marked noise reduction and resolution enhancement in the DLR images. (g) Depicts the relative edge sharpness, measured as the ratio of the maximum gradient in the line profile in the DLR image to that in the original image. A relative edge sharpness of approximately 2.5 was achieved in all three encoding directions, regardless of the applied noise reduction factor. MIP, maximum intensity projection; DLR, deep learning reconstruction.}
    \label{fig:super_resolution}
\end{figure}

Figure~\ref{fig:super_resolution} displays both the original images and the DLR images with a noise reduction factor of 2.5, along with a graph detailing the relative edge sharpness achieved in each encoding direction at various noise reduction factors. A relative edge sharpness of approximately 2.5 was achieved in all three encoding directions, independent of the applied noise reduction factor.

\subsection{Reduction of low spatial resolution artifacts}
The proposed model was trained on pairs of low and high spatial resolution images. It is expected not only to boost apparent resolution but also to reduce artifacts that become more prominent at lower resolutions, such as truncation and intravoxel dephasing artifacts.

To evaluate the model's ability to mitigate these specific types of artifacts, we obtained images where these artifacts are clearly visible and applied DLR to these images, followed by a visual comparison. A brain contrast-enhanced T1-weighted image (CE T1WI) was acquired using a 3D BRAVO pulse sequence, and a susceptibility-weighted image (SWI) was obtained using a 3D SWAN pulse sequence, both on a 3.0T MR scanner (SIGNA Premier, GE Healthcare) with a 48-channel head coil. These images were DL-reconstructed with a noise reduction factor of 2.0.

\begin{figure}[htb!]
    \centering
    \includegraphics[width=1\linewidth]{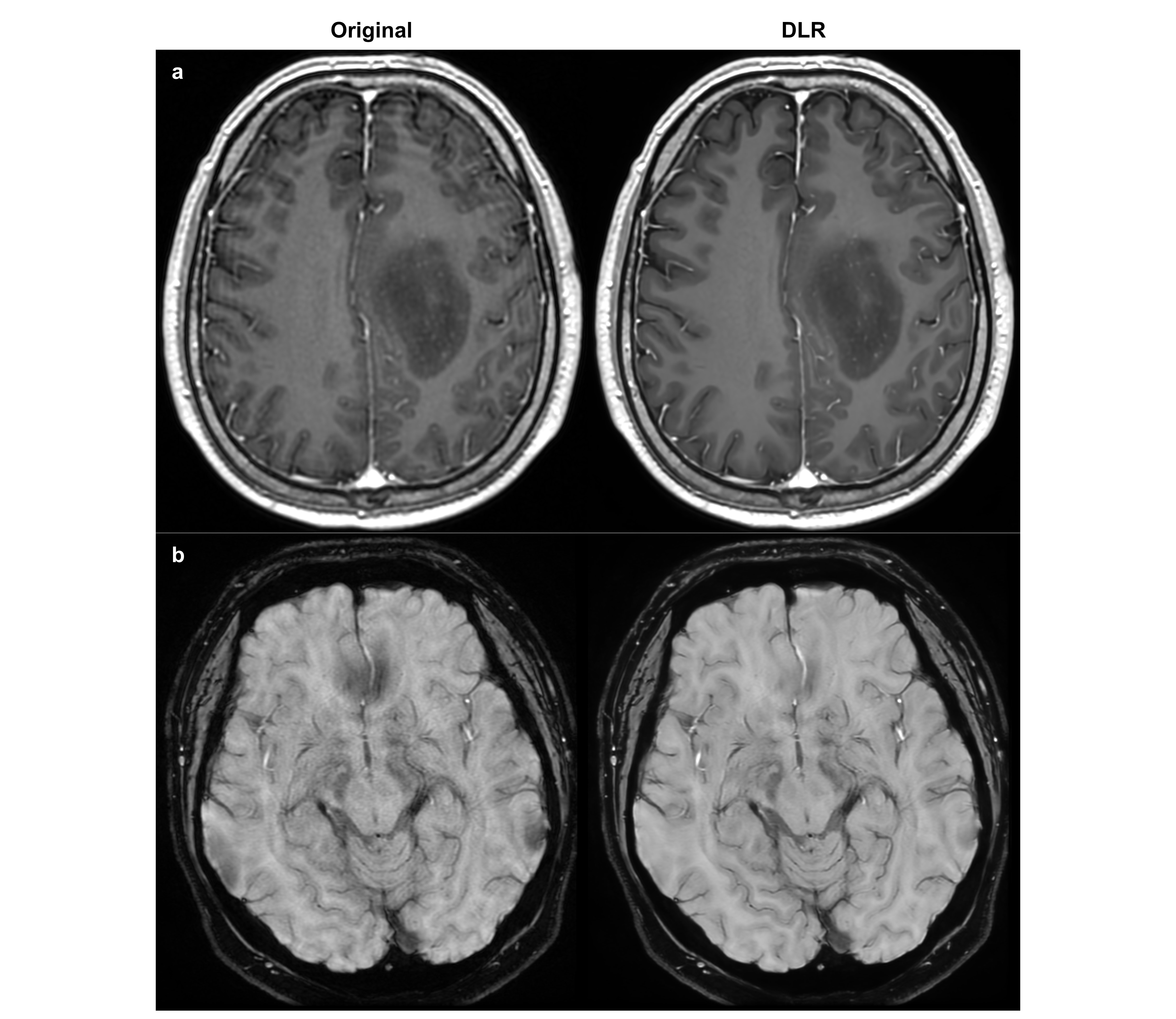}
    \caption{Visual comparison of low spatial resolution artifacts between original and DLR images. (a) Displays truncation artifacts originating from the dura mater of adjacent slices in the original image, with noticeable reduction in the DLR image. (b) Shows intravoxel dephasing artifacts occurring in slices adjacent to the sphenoid sinus, ethmoid air cells, and mastoid air cells in the original image, which are significantly reduced in the DLR image. Improvements in both (a) and (b) are attributed to the slice directional super-resolution. DLR, deep learning reconstruction.}
    \label{fig:low_resolution_artifacts}
\end{figure}

Figure~\ref{fig:low_resolution_artifacts} displays the original images and their DLR images. The DLR CE T1WI shows a significant reduction in truncation artifacts that originated from the dura mater of adjacent slices. This reduction is attributed to the slice directional super-resolution. The DLR SWI demonstrates a notable reduction in intravoxel dephasing artifacts that appeared in slices adjacent to the sphenoid sinus, ethmoid air cells, and mastoid air cells. This improvement is also attributed to the slice directional super-resolution.

\subsection{Compatibility with various scan parameter combinations}
To assess the compatibility of the proposed model with diverse scan parameters that determine $k$-space sampling and the reconstruction pipeline, we obtained multiple images of the same type using different sets of scan parameters. We then applied DLR to these images and performed a visual comparison between the original and DLR images. For this purpose, eight brain fluid attenuated inversion recovery (FLAIR) images were acquired using a 3D turbo spin echo (TSE) pulse sequence on a 3.0 T MR scanner (Ingenia CX, Philips Healthcare) with a 32-channel head coil. The first four images reflect variations in $k$-space sampling, involving changes in SENSE factor, CS factor, elliptical sampling, acquisition voxel size, and receiver bandwidth. These images were acquired through individual scans. The remaining four images reflect variations in the reconstruction pipeline, including adjustments in in-plane interpolation factor, surface coil intensity normalization, geometric distortion correction, and complex component extraction. These images were obtained from a single scan using the scanner console's retrospective reconstruction feature. A summary of these scan parameters is presented in Table~\ref{tab:scan_parameters_for_compatibility}. All acquired images were DL-reconstructed with a noise reduction factor of 4.0, except for those labeled \textit{CS elliptical} and \textit{$k_{max}$ bandwidth}, which were DL-reconstructed with a factor of 5.0.

\begin{table}[htb!]
\renewcommand{\arraystretch}{1.5}
 \caption{MR scan parameter variations for compatibility assessment. }
    \centering
    \scriptsize 
    \begin{tabular}{lccccccccc}
    \toprule
         & \multicolumn{4}{c}{\textbf{$k$-space sampling variations}} & \multicolumn{4}{c}{\textbf{Reconstruction variations}}   \\
         & SENSE  & \makecell[c]{CS \\ elliptical} & \makecell[c]{$k_{max}$  \\ bandwidth} & \makecell[c]{$k_{max}$  \\ bandwidth} & Interpolation  & Intensity  & Distortion  & PSIR \\
       \midrule
         SENSE factor & 2.5 x 2 & No & No & No & 2.5 x 2  & 2.5 x 2 & 2.5 x 2 & 2.5 x 2 \\
         CS factor & No  & \textbf{8} & 7 & 8 & No & No & No & No \\
         \makecell[l]{Elliptical \\ sampling} & No & \textbf{Yes} & No & No & No & No & No & No\\
         \makecell[l]{Voxel size\textsuperscript{*} \\ (mm)} & 0.6 x 0.7  & 0.6 x 0.7 & \textbf{0.8 x 0.9}  & \textbf{1.0 x 1.2} & 0.6 x 0.7 & 0.6 x 0.7  & 0.6 x 0.7  & 0.6 x 0.7 \\
         \makecell[l]{Slice thickness\textsuperscript{*} \\ (mm)} & 2.0 & 2.0  & 2.0  & 2.0 & 2.0 & 2.0  & 2.0  & 2.0 \\         
         \makecell[l]{Bandwidth \\ (Hz/pixel)} & 754.8 & 754.8  & \textbf{951.3}  & \textbf{1291.3} & 754.8 & 754.8  & 754.8  & 754.8 \\
         \makecell[l]{Echo train \\ length} & 150 & 150   & 180  & 215 & 150 & 150  & 150  & 150 \\
         \midrule
         \makecell[l]{In-plane \\ interpolation} & 2.0 & 2.0 & 2.0 & 2.0 & \textbf{1.0} & 2.0 & 2.0 & 2.0 \\
         \makecell[l]{Slice \\ interpolation} & 2.0 & 2.0 & 2.0 & 2.0 & 2.0 & 2.0 & 2.0 & 2.0 \\
         \makecell[l]{Intensity \\ normalization}  & CLEAR & CLEAR & CLEAR & CLEAR & CLEAR & \makecell[c]{\textbf{Body} \\ \textbf{tuned}}  & CLEAR & CLEAR \\
         \makecell[l]{Distortion \\ correction}  & 3D & 3D & 3D & 3D & 3D & 3D & \textbf{No} & 3D \\
         \makecell[l]{Complex \\ component}  & Mag & Mag  & Mag  & Mag & Mag & Mag & Mag & \textbf{Real} \\
       \midrule
         Scan time (s) & 240 & 182 & 115 & 67 & 240  & 240 & 240 & 240 \\
     \bottomrule
     \multicolumn{9}{l}{ \makecell[l]{*refer to the acquisition voxel size, not the interpolated voxel size. SENSE, sensitivity encoding; CS, compressed sensing;\\CLEAR, constant level appearance; PSIR, phase-sensitive inversion recovery; Mag, magnitude.}}
    \end{tabular}
    \label{tab:scan_parameters_for_compatibility}

\end{table}

\begin{figure}[htb!]
    \centering
    \includegraphics[width=1\linewidth]{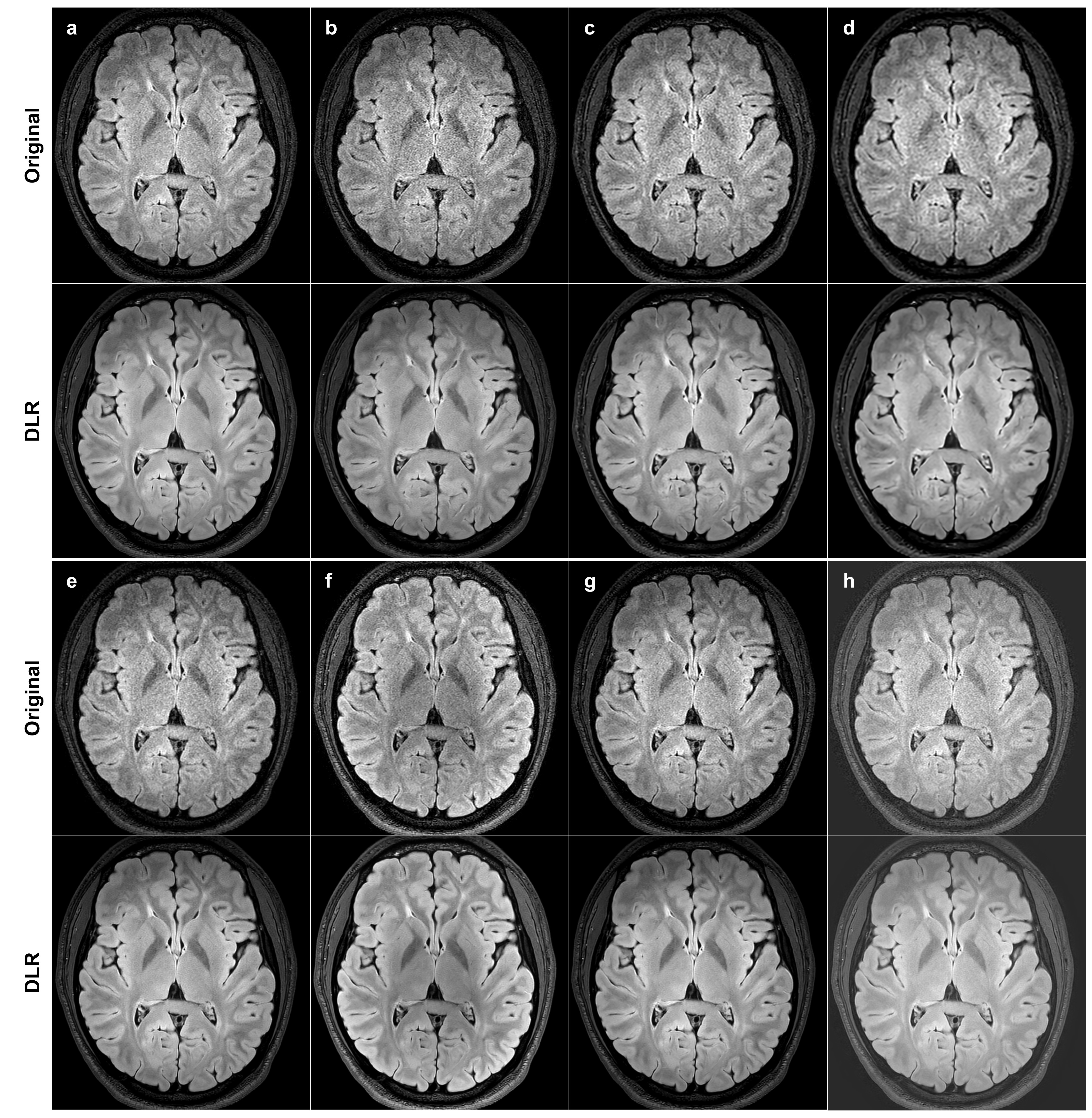}
    \caption{Visual comparison between original and DLR images under various scan parameter settings. (a-d) Demonstrates variations in $k$-space sampling: (a) SENSE (240s), (b) CS and elliptical sampling (182s), (c) low resolution with high receiver bandwidth (115s), and (d) even lower resolution with even higher bandwidth (67s). (e-h) Displays reconstruction pipeline variations, derived from a single scan (a) using the scanner console's delayed reconstruction feature: (e) without in-plane interpolation, (f) with adjusted surface coil intensity normalization, (g) without geometric distortion correction, and (h) PSIR from real component extraction. DLRs across all samples significantly reduce noise and improve overall image quality, while preserving periventricular white matter hyperintensity. SENSE, sensitivity encoding; CS, compressed sensing; PSIR, phase-sensitive inversion recovery; DLR, deep learning reconstruction.
}
    \label{fig:scan_parameter_compatibility}
\end{figure}

Figure~\ref{fig:scan_parameter_compatibility} presents the original images under eight different scan parameter settings and their DLR images. The DLR images consistently exhibit lower noise and improved overall image quality compared to the original images. Notably, periventricular white matter hyperintensity remains unaffected by the DLR.

\subsection{Generalizability across unseen vendors}
To evaluate the model's generalizability beyond the vendors included in the training dataset, we obtained images from unseen vendors and applied DLR to these images, followed by a visual comparison. We focused on three distinct vendors: Canon Medical Systems (Tochigi, Japan), Fujifilm Healthcare (Tokyo, Japan), United Imaging Healthcare (Shanghai, China). An abdomen T2WI was acquired using a 2D fast asymmetric spin-echo (FASE) pulse sequence on a 1.5T MR scanner (Vintage Elan, Canon Medical Systems), a shoulder T2WI with fat suppression was acquired using a 2D fast spin echo (FSE) pulse sequence on a 1.5T MR scanner (Echelon Oval, Fujifilm Healthcare), and a lumbar spine T1WI was acquired using a 2D FSE pulse sequence on a 1.5T MR scanner (uMR 570, United Imaging Healthcare). These images were DL-reconstructed with a noise reduction factor of 2.5.

\begin{figure}[htb!]
    \centering
    \includegraphics[width=1\linewidth]{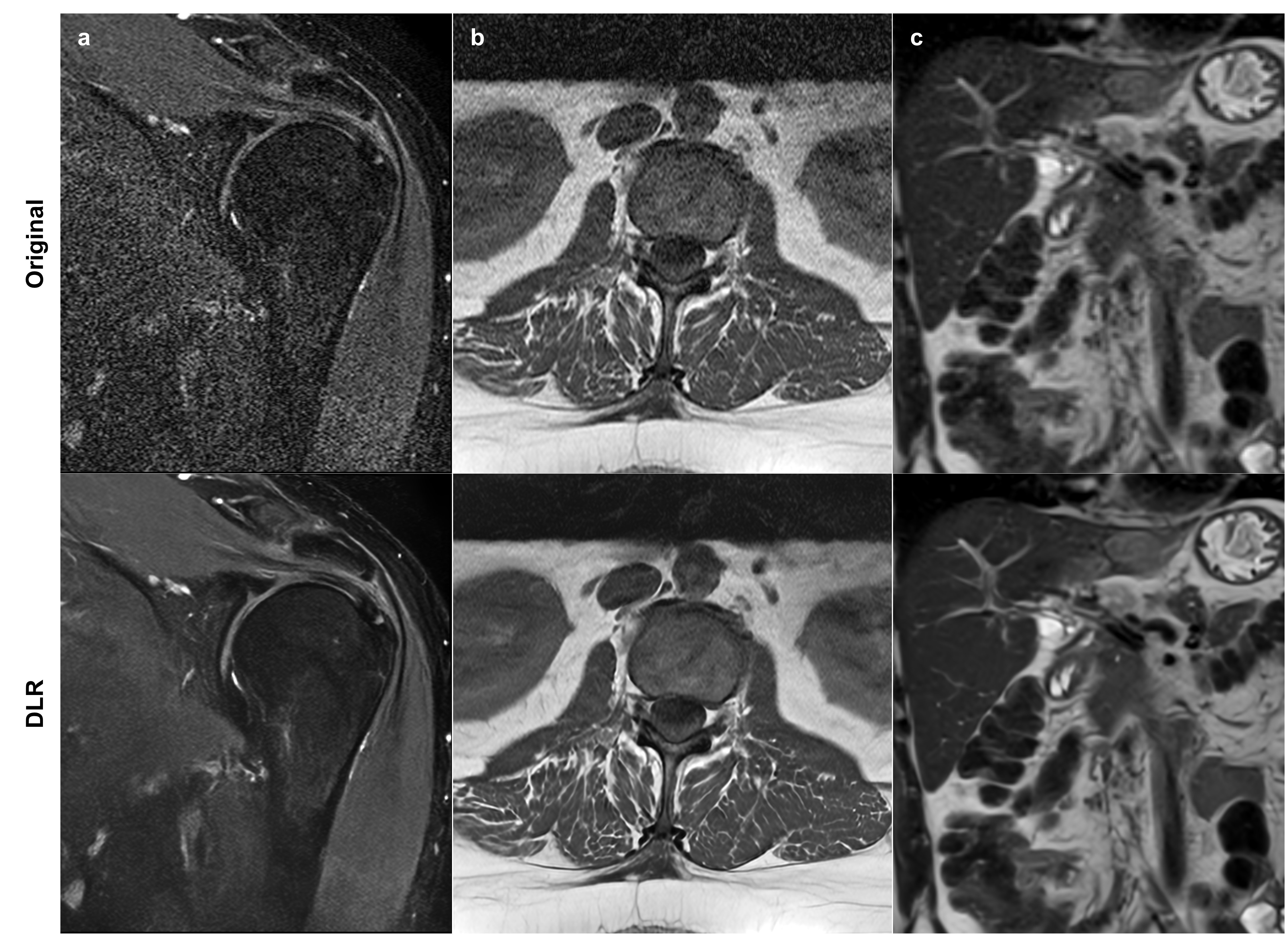}
    \caption{Visual comparison between original and DLR images from vendors not included in the training dataset. (a) Fujifilm Healthcare, (b) United Imaging Healthcare, (c) Canon Medical Systems. In all cases, the DLR images exhibit significant noise reduction and boosted perceived resolution. DLR, deep learning reconstruction.
}
    \label{fig:unseen_vendors}
\end{figure}

Figure~\ref{fig:unseen_vendors} displays the original images and their DLR images from three different unseen vendors. Across all vendors, the DLR images show a reduction in noise and a boost in perceived resolution. Notably, the model demonstrates adaptability to spatially varying noise.

\subsection{Scan time reduction across anatomical regions}
The proposed model is expected to potentially address one of the most common clinical needs, reducing scan time, through a strategic combination with appropriate protocol optimization. This approach involves optimizing the scanning protocol towards reducing scan time, even at the expense of image quality along dimensions where the proposed model is capable of improving.

To explore the model's potential in reducing scan time, we acquired image pairs from standard and accelerated protocols across four different anatomical regions: heart, lumbosacral plexus, knee, and brain, on a 3.0T MR scanner (MAGNETOM Skyra, Siemens Healthineers) with 30-channel body coil/32-channel spine coil, a 32-channel spine coil, a 15-channel transmit/receive knee coil, and a 64-channel head/neck coil, respectively. The accelerated protocols were optimized from standard protocols to reduce scan time and increase spatial resolution, which resulted in increased noise. Table~\ref{tab:scan_parameters_for_scan_time_reduction} summarizes the scan parameters and scan times for each protocol. After applying DLRs to the accelerated images, we visually compared them with the corresponding standard images. A noise reduction factor of 2.0 was used for the heart and knee, while a factor of 3.0 was used for the lumbosacral plexus and brain. For the knee T2WIs, a 1 mm thick multiplanar reformation (MPR) image was generated from each source image (standard, accelerated, and accelerated DLR).

\begin{table}[htb!]
\renewcommand{\arraystretch}{1.5}
 \caption{MR scan parameters for standard and accelerated protocols across different anatomical regions.}
    \centering
    \scriptsize 
    \begin{tabular}{lccccccccc}
    \toprule
         & \multicolumn{2}{c}{Heart} & \multicolumn{2}{c}{Lumbosacral plexus} & \multicolumn{2}{c}{Knee} & \multicolumn{2}{c}{Brain}   \\
         & Standard  & Fast & Standard & Fast & Standard  & Fast  & Standard  & Fast \\
       \midrule
         Image type & \multicolumn{2}{c}{Cine} & \multicolumn{2}{c}{Neurography} & \multicolumn{2}{c}{T2WI} & \multicolumn{2}{c}{DWI} \\
         Pulse sequence & \multicolumn{2}{c}{2D TRUFI} & \multicolumn{2}{c}{3D SPACE} & \multicolumn{2}{c}{3D SPACE} & \multicolumn{2}{c}{2D EPI} \\
         \makecell[l]{Acceleration \\ factor} & \makecell[c]{GRAPPA \\ 2} & \makecell[c]{GRAPPA \\ 3} & \makecell[c]{GRAPPA \\ 2 x 1} & \makecell[c]{GRAPPA \\ 6 x 1} & \makecell[c]{CAIPI \\ 1 x 3}  & \makecell[c]{CAIPI \\ 2 x 3} & \makecell[c]{GRAPPA \\ 2} & \makecell[c]{GRAPPA \\ 3} \\
         \makecell[l]{Phase \\ resolution (\%)} & 70  & 80 & 80 & 100 & 85 & 95 & 100 & 100 \\
         \makecell[l]{Slice \\ resolution (\%)} & - & - & 50 & 50 & 75 & 83 & - & - \\
         \makecell[l]{Phase \\ partial Fourier} & Off  & Off & Allowed & Allowed & Allowed & Allowed & 6/8 & 7/8 \\     
         \makecell[l]{Slice \\ partial Fourier} & -  & - & Off & Off & Off & Off & - & - \\     
         \makecell[l]{Phase \\ oversampling (\%)} & 80  & 100 & 50  & 50 & 0  & 0  & 0 & 0 \\
         \makecell[l]{Slice \\ oversampling (\%)} & - & -  & 27.3  & 27.3 & 20 & 10  & -  & - \\         
         TSE turbo factor & - & -  & 128  & 110 & 65 & 65  & -  & - \\
         \makecell[l]{Diffusion b1000 \\ averages} & - & -  & -  & - & - & -  & 2  & 1 \\
       \midrule
         Scan time (s) & 12\textsuperscript{*} & 9\textsuperscript{*} & 369 & 219 & 321  & 185 & 52 & 32 \\
     \bottomrule
     \multicolumn{9}{l}{ \makecell[l]{*Breath-hold time. Each slice was scanned in a single breath-hold. \\ T2WI, T2-weighted image; DWI, diffusion-weighted image; TRUFI, true fast imaging with steady-state free precession;\\ SPACE, sampling perfection with application optimized contrast using different flip angle evolutions; EPI, echo planar \\imaging;  GRAPPA, GeneRalized Autocalibrating Partially Parallel Acquisitions; CAIPI, Controlled Aliasing in Parallel \\ Imaging Results in Higher Acceleration; TSE, turbo spin echo.}}
    \end{tabular}
    \label{tab:scan_parameters_for_scan_time_reduction}

\end{table}

Figure~\ref{fig:scan_time_reduction} presents standard, accelerated, and accelerated DLR images across four anatomical regions. The DLR heart cine image shows improved delineation of papillary muscles in the left ventricle. The DLR lumbosacral plexus neurography image provides clearer visualizations of nerve roots, dorsal root ganglia, and spinal nerves. The DLR knee T2WI reveals a more distinct depiction of the overall soft tissues. The DLR brain DWI exhibits superior apparent resolution, particularly in the delineation of sulci. Notably, it also demonstrates reductions in geometric distortion artifacts in the frontal lobe, attributed to the increased GRAPPA factor. In all cases, the accelerated DLR images exhibit equivalent or reduced noise levels compared to the standard images.

\begin{figure}[htb!]
    \centering
    \includegraphics[width=1\linewidth]{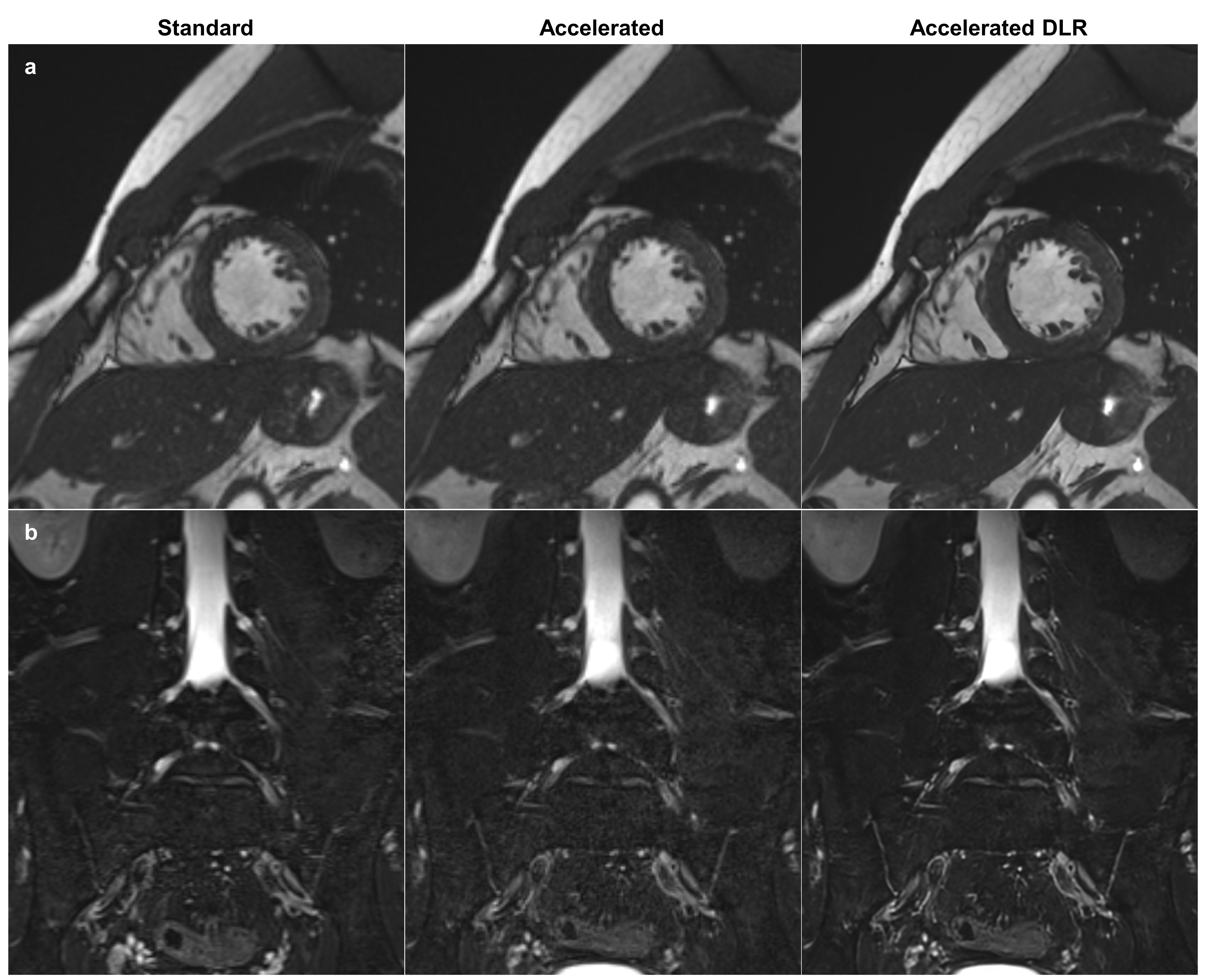}
    \caption{Visual comparison between standard, accelerated, and accelerated DLR images across different anatomical regions. The accelerated protocols were optimized from standard protocols to reduce scan time and increase spatial resolution, which resulted in increased noise. (a) Heart cine images where each slice was acquired in a single breath-hold, with breath-hold times of 12 seconds for standard and 9 seconds for accelerated scans. (b) Lumbosacral plexus neurography images with scan times of 369 seconds for standard and 219 seconds for accelerated. (c) Reformatted knee T2-weighted images with scan times of 321 seconds for standard and 185 seconds for accelerated. (d) Brain diffusion-weighted images with scan times of 52 seconds for standard and 32 seconds for accelerated. In all cases, the accelerated DLR images exhibit boosted perceived resolution and either equivalent or reduced noise levels compared to the standard images. DLR, deep learning reconstruction.
    }
    \label{fig:scan_time_reduction}
\end{figure}

\begin{figure}[htb!]
\ContinuedFloat
    \centering
    \includegraphics[width=1\linewidth]{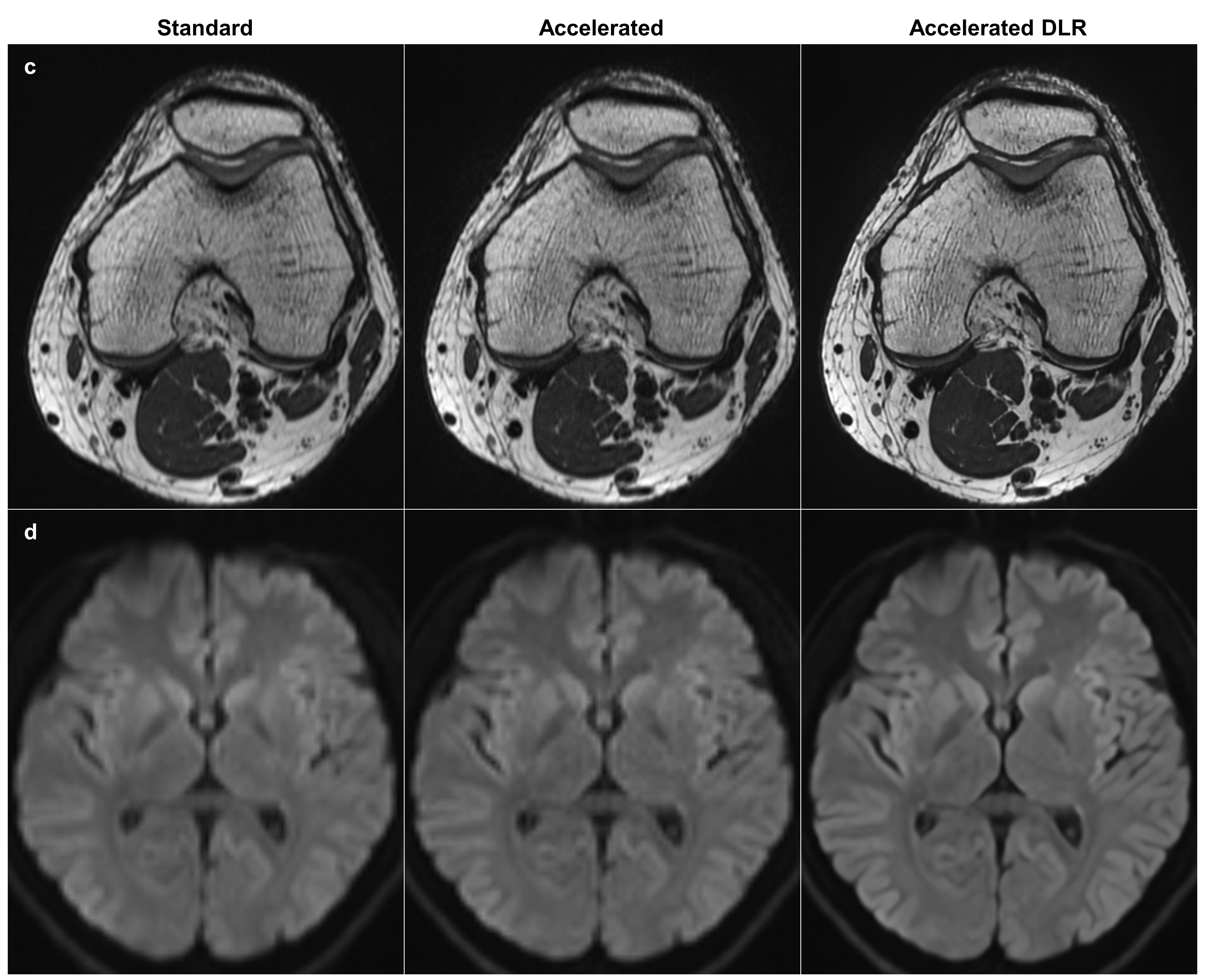}
    \caption{Visual comparison between standard, accelerated, and accelerated DLR images across different anatomical regions (continued).}
    \label{fig:scan_time_reduction2}
\end{figure}

\section{Discussion}
This work introduces a DICOM-based, \textit{all-in-one} DL framework designed to enhance MR image quality across multiple aspects of $k$-space sampling and provide comprehensive coverage across various imaging scenarios. The model demonstrated multi-dimensional image improvements, tunable denoising with high accuracy, effective super-resolution in all encoding directions, and reductions of low spatial resolution artifacts. It also showed compatibility with various scan parameter combinations and generalizability across unseen vendors. The model also facilitated scan time reductions across anatomical regions while surpassing standard image quality, showcasing its ability to break the conventional trade-off between scan time and image quality. Further evaluation is needed to fully assess the model's compatibility with even more varied conditions and to explore its limits in reducing scan times by optimizing multiple dimensions.

The proposed framework for generating training pairs enables the production of virtually unlimited numbers of training pairs from a single raw $k$-space dataset by simulating various scan parameter scenarios. This approach not only serves as a form of data augmentation but also provides a more accurate representation of actual scanner images compared to traditional DL augmentation techniques such as rotation and flipping.

The learning task varies for each training pair due to the diversity of simulated sampling scenarios. In previous approaches, often exemplified by settings where the target is defined as fully-sampled data and the input as uniformly undersampled data by a factor of 4, the learning task remains consistent across all training pairs. However, in the proposed framework, the relationship between input and target changes with each pair, suggesting that traditional single models might not be ideally suited for such varied tasks. Our model, which leverages the CE U-Net to integrate contextual data through a dynamic modulation pathway, appears to be effectively adapted to these diverse learning tasks.

The proposed model distinguishes itself by utilizing conventionally reconstructed image data as inputs, rather than raw $k$-space data. This approach's primary advantage is that it enables the model to operate on a DICOM-based framework. Such a capability significantly enhances the model's clinical applicability, as DICOM data are widely available and commonly possessed by clinical centers, unlike the less accessible raw $k$-space data. Furthermore, this strategy offers computational benefits, including faster processing speeds and lower GPU memory demands. An additional merit includes the potential for greater generalizability across coil configurations compared to models that take raw $k$-space data as inputs. This advantage has likely influenced the demonstrated generalizability of the proposed model to unseen vendors. A notable challenge for DICOM-based models, compared to raw $k$-space based models, is their adaptability to spatially varying noise due to geometric distortion correction and surface coil intensity normalization \cite{lebel2020performance}. The proposed framework has effectively addressed this challenge by incorporating simulations of these scenarios, thereby demonstrating adaptability in these areas.

All performance evaluations in this work, including quantitative assessments, were conducted on images of human subjects rather than on phantom images. Phantom images have the advantage of simpler structures, including homogeneous areas and abrupt edges, which make them convenient for measuring noise level and edge sharpness. However, the performance on phantom images does not necessarily guarantee similar results on clinical human images, and this is particularly true for DL-based models where performance can be significantly influenced by the type of input image. The proposed model has demonstrated its performance on human images, underscoring its valuable potential in clinical applications.

This work's performance evaluation faces two main limitations: first, the absence of image quality and diagnostic performance assessments by board-certified radiologists, and second, the inability to conduct statistical analysis due to the limited number of volunteers. However, the clinical value of this model has been substantiated through several clinical validation studies. These studies involved comprehensive evaluations by board-certified radiologists and statistical analyses supported by adequate patient numbers, offering a potential offset to the noted limitations. Yoo et al. \cite{yoo2023deep} demonstrated the model could reduce scan times by an average of 32.3\% without compromising image quality or diagnostic performance in degenerative lumbar spine diseases, based on a study involving four radiologists and 50 patients. Lee et al. \cite{lee2023highly} conducted a multi-vendor study with three radiologists assessing 45 patients, showing that the model could achieve comparable image quality and diagnostic performance while reducing scan times by an average of 41.0\% in knee MRI. Suh et al. \cite{suh2024improving} showcased in their study, which involved two radiologists evaluating 117 patients, that applying this model to a thinner slice thickness protocol could enhance diagnostic performance and image quality for temporal lobe epilepsy evaluation, compared to a routine protocol without the model.

\section{Conclusions}
The proposed \textit{all-in-one} DL framework enables a single model to enhance MR image quality in a multi-dimensional manner and to be compatible across a broad spectrum of scenarios, including various vendors, field strengths, pulse sequences, contrast weightings, scan parameters, and anatomical regions. Its DICOM-based operation particularly enhances its applicability for real-world applications. Serving as the core algorithm of the commercially available product SwiftMR\textsuperscript{TM}, the proposed model has been effectively utilized in numerous clinical applications to improve image quality and reduce scan times. Given its demonstrated effectiveness and versatility, we expect its use to expand in the field of clinical MRI.

\section*{Acknowledgments}
We thank Sohyun Kim of AIRS Medical for her support and encouragement.

\section*{Conflicts of Interest}
All co-authors are employees of AIRS Medical, except Jeewook Kim, a former employee.

\clearpage

{\small
\bibliographystyle{plainnat}
\bibliography{SwiftMR}
}
       
\end{document}